\pdfoutput=1
\documentclass[prx,twocolumn,floatfix,superscriptaddress,aps,longbibliography]{revtex4-1}
\usepackage{amsfonts,amsmath,amssymb,color,times,graphicx}
\definecolor{darkblue}{rgb}{0, 0, 0.8}
\usepackage[colorlinks=true, breaklinks=true, linkcolor=blue, citecolor=darkblue, urlcolor=darkblue]{hyperref}
\usepackage{graphicx}
\usepackage{bmpsize}
\usepackage{amsfonts}
\usepackage{gensymb}
\usepackage{bm}

\newcommand{\ccs}{CsCeSe$_2$}
\newcommand{\mb}{$\mu_{\rm B}$}

\begin{document}
\title{Stripe magnetic order and field-induced quantum criticality in the perfect triangular-lattice antiferromagnet \ccs}
\author{Tao Xie}
\thanks{Corresponding author: xiet69@mail.sysu.edu.cn}
\affiliation{Center for Neutron Science and Technology, Guangdong Provincial Key Laboratory of Magnetoelectric Physics and Devices, School of Physics, Sun Yat-sen University, Guangzhou, Guangdong 510275, China}
\affiliation{Neutron Scattering Division, Oak Ridge National Laboratory, Oak Ridge, Tennessee 37831, USA}
\author{N. Zhao}
\affiliation{Department of Physics, Southern University of Science and Technology, Shenzhen, Guangdong, 518055, China.}
\author{S. Gozel}
\affiliation{Laboratory for Theoretical and Computational Physics, Paul Scherrer Institute, CH-5232 Villigen-PSI, Switzerland}
\author{Jie Xing}
\affiliation{Materials Science and Technology Division, Oak Ridge National Laboratory, Oak Ridge, Tennessee 37831, USA}
\author{S. M. Avdoshenko}
\affiliation{Leibniz-Institut f\"ur Festk\"orper- und Werkstoffforschung (IFW Dresden), Helmholtzstra{\ss}e 20, 01069
Dresden, Germany}
\author{K. M. Taddei}
\affiliation{Neutron Scattering Division, Oak Ridge National Laboratory, Oak Ridge, Tennessee 37831, USA}
\affiliation{Advanced Photon Source, Argonne National Laboratory, Lemont, Illinois 60439, USA}
\author{A. I. Kolesnikov}
\affiliation{Neutron Scattering Division, Oak Ridge National Laboratory, Oak Ridge, Tennessee 37831, USA}
\author{L.~D.~Sanjeewa}
\affiliation{Materials Science and Technology Division, Oak Ridge National Laboratory, Oak Ridge, Tennessee 37831, USA}
\author{Peiyue~Ma}
\affiliation{Center for Neutron Science and Technology, Guangdong Provincial Key Laboratory of Magnetoelectric Physics and Devices, School of Physics, Sun Yat-sen University, Guangzhou, Guangdong 510275, China}
\author{N.~Harrison}
\affiliation{National High Magnetic Field Laboratory, Los Alamos National Laboratory, Los Alamos, New Mexico 87545, USA}
\author{C. dela Cruz}
\affiliation{Neutron Scattering Division, Oak Ridge National Laboratory, Oak Ridge, Tennessee 37831, USA}
\author{L. Wu}
\affiliation{Department of Physics \&\ Academy for Advanced Interdisciplinary Studies, Southern University of Science \&\ Technology, Shenzhen, Guangdong, 518055, China}
\author{Athena~S.~Sefat}
\affiliation{Materials Science and Technology Division, Oak Ridge National Laboratory, Oak Ridge, Tennessee 37831, USA}
\author{A. L. Chernyshev}
\affiliation{Department of Physics and Astronomy, University of California, Irvine, California 92697, USA}
\author{A. M. L\"auchli}
\affiliation{Laboratory for Theoretical and Computational Physics, Paul Scherrer Institute, CH-5232 Villigen-PSI, Switzerland}
\affiliation{Institute of Physics, Ecole Polytechnique F\'ed\'erale de Lausanne (EPFL), CH-1015 Lausanne, Switzerland}
\author{A.~Podlesnyak}
\affiliation{Neutron Scattering Division, Oak Ridge National Laboratory, Oak Ridge, Tennessee 37831, USA}
\author{S.~E.~Nikitin}
\thanks{Corresponding author: stanislav.nikitin@psi.ch}
\affiliation{Laboratory for Neutron Scattering and Imaging, Paul Scherrer Institut, CH-5232 Villigen-PSI, Switzerland}

\begin{abstract}
The two-dimensional triangular-lattice antiferromagnet (TLAF) is a textbook example of frustrated magnetic systems. Despite its simplicity, the TLAF model exhibits a highly rich and complex magnetic phase diagram, featuring numerous distinct ground states that can be stabilized through frustrated next-nearest-neighbor couplings or anisotropy. In this paper, we report low-temperature magnetic properties of the TLAF material \ccs. The inelastic neutron scattering (INS) together with specific heat measurements and density functional theory calculations of crystalline electric field suggest that the ground state of Ce ions is a Kramers doublet with strong easy-plane anisotropy. Elastic neutron scattering measurements demonstrate the presence of stripe-$yz$ magnetic order that develops below $T_{\rm N} = 0.35$~K, with the zero-field ordered moment of $m_{\rm Ce} \approx 0.65~\mu_{\rm B}$. Application of magnetic field first increases the ordering temperature by about 20\%\ at the intermediate field region and eventually suppresses the stripe order in favor of the field-polarized ferromagnetic state via a continuous quantum phase transition (QPT). The field-induced response demonstrates sizable anisotropy for different in-plane directions, $\mathbf{B}\parallel{}\mathbf{a}$ and $\mathbf{B}\perp{}\mathbf{a}$, which indicates the presence of bond-dependent coupling in the spin Hamiltonian. We further show theoretically that the presence of anisotropic bond-dependent interactions can change the universality class of QPT for $\mathbf{B}\parallel{}\mathbf{a}$ and $\mathbf{B}\perp{}\mathbf{a}$.
\end{abstract}

\maketitle

\section{Introduction}\label{sec:intro}

\begin{figure}[t]
\center{\includegraphics[width=1\linewidth]{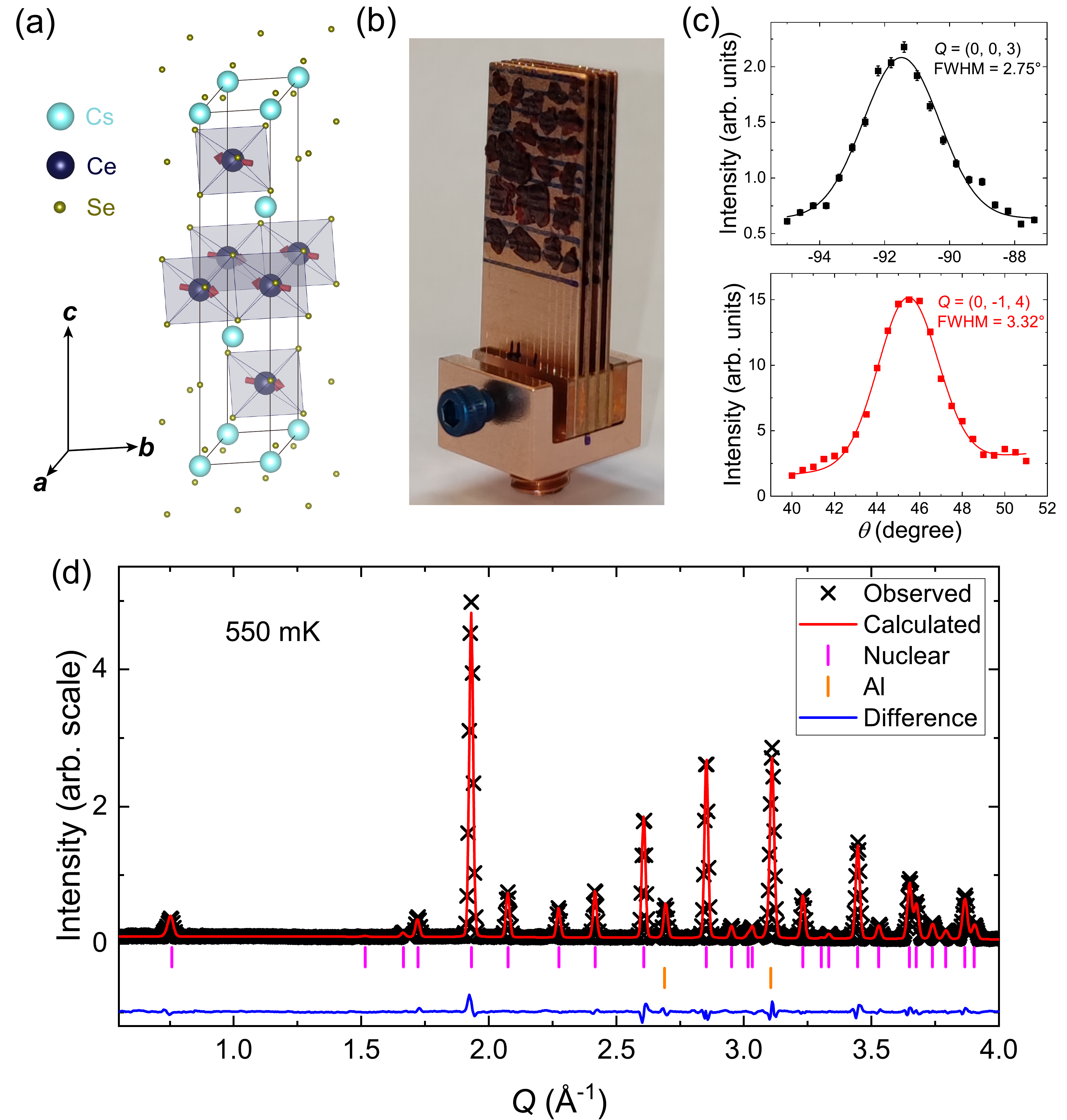}}
  \caption{
  (a)~Crystal structure of \ccs. The light cyan, navy, and dark yellow balls represent Cs, Ce, and Se atoms respectively. The red arrows indicate the ordered moments of the stripe-$yz$ AFM order at zero field.
  (b)~The sample used for the single crystal neutron measurements.
  (c)~Rocking scans at (0,~0,~3) and (0,~$-$1,~4) Bragg peaks demonstrate good mosaicity of the sample.
  (d)~Neutron powder diffraction pattern collected above $T_{\rm N}$ shows good quality of the powder sample under study.
  }
  \label{Fig:intro}
\end{figure}

The family of materials known as delafossites, with a general chemical composition of $ABX_2$, has emerged as a key area of interest in modern solid-state physics because of unusual and versatile physical behaviors~\cite{mackenzie2017properties}. The $A$ site in this structure is occupied by monovalent cations such as alkalis, Pd or similar; trivalent transition- or rare-earth cations occupy the $B$ site and $X$ is two-valent anion such as O, S or Se. The common structural motif is triangular layers of $BX_2$ that are stacked along $c$ axis in a different fashion~[Fig.~\ref{Fig:intro} (a)]. Depending on the composition, these materials can exhibit either insulating or metallic behavior. Certain metallic materials, such as PdCoO$_2$ and PdCrO$_2$, are particularly noteworthy because of very high electrical conductivity resulting from their exceptional purity, which enables the observation of mesoscopic transport effects~\cite{bachmann2019super, mcguinness2021low} and Planckian scattering~\cite{zhakina2023investigation}.

On the other hand, insulating rare-earth-based materials are considered as a promising platform for the investigation of frustrated magnetism~\cite{liu2018rare, schmidt2021yb}. In particular, Yb-based materials NaYbO$_2$ and NaYbSe$_2$ were proposed to host the quantum spin-liquid (QSL) state~\cite{ding2019gapless, Ranjith2019naybse, bordelon2019field, dai2020spinon}. Two other Yb-based delafossites, CeYbSe$_2$ and KYbSe$_2$, exhibit 120$^{\circ}$ ordered ground state~\cite{xie2023complete,Scheie2021}, and lie in the proximity of the QSL phase because of competition between first and second nearest-neighbor interactions. However, Ce- and Er-based delafossites, KCeS$_2$~\cite{Bastie2020, Kulbakov2021} and KErSe$_2$~\cite{ding2023stripe}, exhibit stripe order of the rare-earth magnetic moments, highlighting the difference with Yb-based sister materials, despite the same lattice structure within the $ab$ plane and $S = 1/2$ magnetism. Moreover, thermodynamic measurements of KCeS$_2$ have shown that its phase diagram is anisotropic for the magnetic field applied along different in-plane directions, which is considered as a manifestation of the bond-dependent interactions~\cite{Bastie2020}. Altogether, Ce-based delafossite is a promising class of materials for experimental study of the effect of bond-dependent interactions on the low-energy physics of the triangular-lattice antiferromagnet (TLAF).

In this paper, we apply powder and single-crystal neutron diffraction, thermodynamic probes, exact diagonalization (ED), and density functional theory (DFT) calculations to study the magnetic phase diagram and magnetic structure of CsCeSe$_2$. High-energy inelastic neutron scattering (INS) data along with DFT calculations indicate that Ce$^{3+}$ ions have a well-isolated doublet as the ground state. Thermodynamic measurements demonstrate a long-range magnetic order at zero field below transition temperature $T_{\rm N} = 0.35$~K and reentrant-like phase diagram with clear anisotropy for the in-plane magnetic field applied along and perpendicular to the $a$ axis. The saturation fields along $a$ axis, perpendicular to $a$ axis, and along $c$ axis are $B_{\mathrm{sat}}^{\| a}\approx{}3.86$~T, $B_{\mathrm{sat}}^{\bot a}\approx{}3.52$~T, and $B_{\mathrm{sat}}^{\| c}\approx{}13.5$~T, respectively. Neutron diffraction data confirm that the zero-field magnetic structure is a stripe-$yz$ order with the propagation wavevector $\mathbf{k} = (0,~1/2,~1)$ and ordered moment $m_{\rm Ce} \approx{}0.65~\mu_{\rm B}$. By utilizing ED calculations we demonstrate that the universality class of the field-induced quantum phase transition (QPT) in \ccs\
is 2+1D Ising at $T$ = 0 when considering isolated layers with magnetic field along the $a$ axis. At finite temperatures, the universality class changes to a simple 2D Ising case. We also discuss the nature of the QPT for magnetic field perpendicular to the $a$ axis, and when the inter-layer coupling is considered.

\section{Experimental details}\label{sec:exp}
We prepared high-quality single crystals of \ccs~ using a flux method~\cite{Xing20191}. The chemical composition measurements were carried out using femtosecond laser-ablation-inductively coupled plasma time-of-flight mass spectrometry (LA-ICP-TOF) and energy-dispersive x-ray spectroscopy (EDS).
The low-temperature magnetization measurements were performed on a Hall sensor magnetometer integrated with a dilution refrigerator (DR) insert~\cite{QDMagnetometry,Cavallini2004,Candini2006}. The obtained Hall resistance is proportional to the magnetization. Additional magnetization measurements  were performed using MPMS-VSM at $T \geq\ 2$~K to normalize the Hall resistance to absolute units of magnetization. The heat capacity data and magnetocaloric effect (MCE) results with temperature down to 60 mK and magnetic field up to 8 T were collected in a PPMS with dilution refrigerator insert. We performed neutron powder diffraction measurements on the HB-2A (POWDER) diffractometer at High Flux Isotope Reactor, Oak Ridge National Laboratory. To achieve the desired temperature, a He-3 refrigerator was used. Data were collected at 250 mK and 550 mK with incident neutron wavelength $\lambda$ = 2.41 {\AA}. For the single-crystal neutron scattering experiments, we coaligned about 90 single crystals in the $(0~K~L)$ scattering plane on thin copper plates to get a mosaic sample with mass $\sim$0.4 g and mosaicity $\sim$3\degree~[Figs.~\ref{Fig:intro}(b) and ~\ref{Fig:intro}(c)]. The single-crystal neutron scattering experiments were performed at the time-of-flight Cold Neutron Chopper Spectrometer (CNCS)~\cite{CNCS1, CNCS2} and Fine-Resolution Fermi Chopper Spectrometer (SEQUOIA)~\cite{SEQ} at the Spallation Neutron Source, Oak Ridge National Laboratory. The data from CNCS were collected with an incident neutron energy $E_{\mathrm{i}}$ = 1.55~meV, providing an energy resolution of 0.05 meV.
During the measurements, the vertical magnetic field was applied along the [$-$1 0.5 0] direction in $ab$ plane, using a 5-T cryomagnet equipped with a dilution refrigerator. The high-energy spin excitations from SEQUOIA were collected with $E_{\mathrm{i}}$ = 100~meV at $T$ = 5 and 150 K. We used the software packages~\textsc{MantidPlot}~\cite{Mantid}, ~\textsc{Horace}~\cite{Horace}, and ~\textsc{Dave}~\cite{Dave} for data reduction and analysis. The magnetic structure refinements were performed with the Rietveld method using $\textsc{FullProf}$~\cite{FullProf1993} and $\textsc{Mag2Pol}$~\cite{qureshi2019mag2pol} software.

\section{Results}\label{sec:res}

\subsection{Single-ion properties}\label{sec:CEF}

\begin{figure}[t]
\center{\includegraphics[width=1\linewidth]{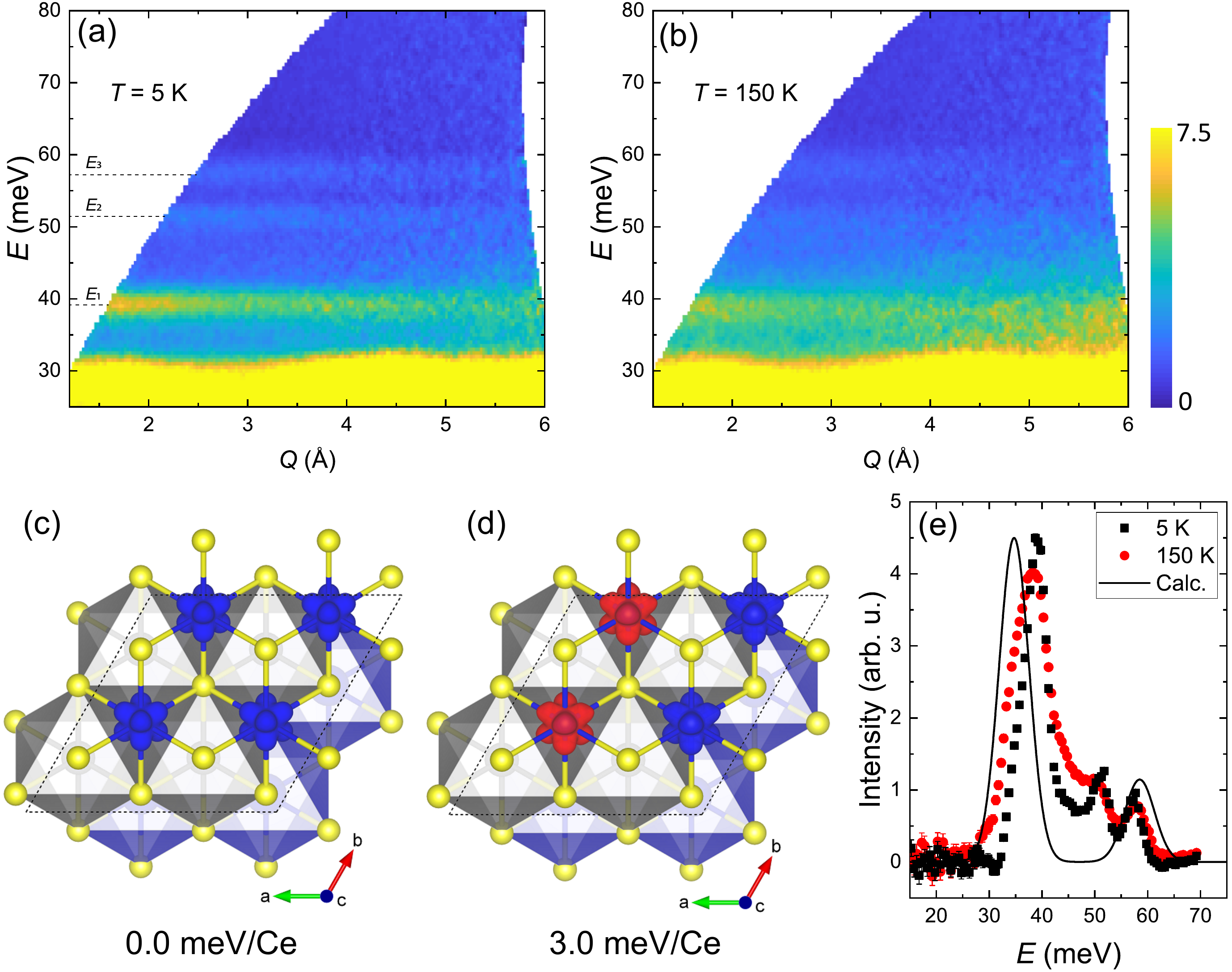}}
  \caption{~Crystal electric field splittings of \ccs.
  (a)~Powder-averaged spectrum measured at $T$ = 5~K. The dashed lines indicate the three excitations at $E_1$, $E_2$, and $E_3$.
  (b)~Powder-averaged spectrum measured at $T$ = 150~K.
  (c)--(d)~The top view of 2$\times$2$\times$1 supercell of \ccs with spin density isosurfaces ($\rho_\alpha -\rho_\beta$) at DFT/PBE/PAW level shown for solutions with FM (c) and AFM (d) spin constraints.
  (e)~1D cuts of the spectra taken at $T$ = 5 and 150~K. The solid line is the CASSCF(1,7)/ANO-RCC-VDZP/RASSI/SO calculation of the CEF at $T$ = 0~K.
  }
  \label{CEF}
\end{figure}

We first present the single-ion properties focusing on the crystal electric field (CEF) splitting. The data were collected at SEQUOIA with $E_{\rm i}$ = 100~meV, and the results are presented in Fig.~\ref{CEF}. The spectrum measured at $T$ = 5 K shows three CEF-like excitations at $E_1$ = 38.70(5) meV, $E_2$ = 50.95(2) meV, and $E_3$ = 57.33(17) meV [see Fig.~\ref{CEF}(a)]. The peaks do not change the positions when increasing temperature to $T$ = 150 K, but become broader with a slight ($<$10$\%$) decrease in the intensity [see Fig.~\ref{CEF}(b) and \ref{CEF}(e)]. Such a decrease is caused by the thermal population of the excited states and is in agreement with simulations based on the CEF Hamiltonian (below) that predict $\approx$8$\%$ intensity decrease for the excitations at 40-50 meV at 150 K.
However, as the ground multiplet of Ce$^{3+}$, $^2F_{5/2}$ can be split into three Kramers doublets, which will give rise to only two excitation peaks in the INS spectrum. This means that one of these peaks has an extrinsic origin.

Following the computational strategy used for the KCeS$_2$ system~\cite{Kulbakov2021}, we employed a similar DFT level of theory in our attempt to estimate mean field exchange interactions and CEF splitting scheme. The DFT/PBE/PAW assessment was performed using the Vienna $ab\ initio$ Simulation Package (VASP) and the projector augmented-wave method as implemented in the v5.5.4 code and the standard pseudopotential~\cite{KresseHafner93, PerdewErnzerhof96}. The experimental geometry was optimized. We did not find any significant changes in geometry after the optimization compared to the experimental value down to the residual force of 10$^{-4}$eV/\AA. The spin-polarized solution for the single unit cell has converged to the high-spin solution. Furthermore, we have used the 2$\times$2$\times$1 supercell to compute the system with ferromagnetic (FM) and antiferromagnetic (AFM) constraints in the CeSe$_2$ layer. Unlike in KCeS$_2$, the FM state remains more stable ($\Delta E$ = 3.0~meV/Ce) for this vanilla DFT model [Figs.~\ref{CEF}(c) and ~\ref{CEF}(d)].

To provide insight into the CEF structure, we used the [CeSe$_6$]$^{3+}$ cluster model, with Se ions substituted by Bader point charges. We computed the full spin-orbit coupling (SOC) electronic structure using CASSCF(1,7)/ANO-RCC-VDZP/RASSI/SO  level theory as implemented in the \textsc{OpenMolcas} code~\cite{AquilanteAutschbach20, ChibotaruUngur12, Rotter04}. With system 4$f^1$ in active space, the reference space of the problem is limited to seven roots. These roots were used in the state interaction problem with the SO-Hamiltonian basis of all 14 SOC states. To achieve a better agreement between the theoretical CEF energies and INS signals, the Bader charges of Se site were scaled by 5$\%$. The calculated $B_l^m$ parameters are: $B_2^0 = 2.57$~meV, $B_4^0 = -0.05$~meV, and $B_4^3 = 1.67$~meV. The two calculated nonzero energy levels of 34.7 and 58.5~meV are close to the observed INS peaks at $E_1$ and $E_3$ and have very similar intensity ratio. It should be noted that the CEF energies 34.7 and 58.5 meV in Table~\ref{Table:Th1} are direct CASSCF calculations, whose mismatches with the experimental results ($E_1$ and $E_3$) are less than the 10\%, which is appropriate for this model. Thus, we conclude that the observed INS peaks at $E_1$ and $E_3$ are the CEF splittings of the $^2F_{5/2}$ multiplet.

\begin{table*}[tb]
\caption{$^2F_{5/2}$ Multiplet structure and wavefunction composition at CASSCF/RASSI/SO level for Ce$^{3+}$ in CeSe$_6$ polyhedron with experimental geometry.}
\centering
\begin{ruledtabular}
\begin{tabular}{ccccccc}
System             & Energy, meV & Wavefunction composition, \% of $|J, m_J\rangle$ & g$_x$ & g$_y$ & g$_z$ & INS, barn  \\
\hline
                             & 0.00       &    0.92 $\vert \pm 1/2 \rangle$ + 0.08 $\vert \mp 5/2 \rangle$    & 2.30     & 2.30     & 0.50     & 1.73      \\
  CsCeSe$_2$                 & 34.70      &    0.79 $\vert \pm 3/2 \rangle$ + 0.21 $\vert \mp 3/2 \rangle$    & 0.00     & 0.00     & 2.70     & 1.73      \\
 (model CeSe$_6$)            & 58.50      &    0.77 $\vert \pm 5/2 \rangle$ + 0.15 $\vert \mp 1/2 \rangle$& 0.25     & 0.25     & 3.86     & 0.44      \\
\end{tabular}
\end{ruledtabular}
\label{Table:Th1}
\end{table*}

The origin of the extra INS peak at $E_2$ is currently unknown and below we exclude some known scenarios in Ce-based compounds. Similar extra local INS peak has been widely observed in the sister compounds KCeO$_2$, KCeS$_2$~\cite{Bastie2020,Bordelon2021,Kulbakov2021}, and other Ce-based compounds such as CeAuAl$_3$~\cite{vcermak2019} and Ce$_2$Zr$_2$O$_7$~\cite{Gaudet2019}, where the chemical-impurity origin, CEF-phonon coupling scenario, intermultiplet-transition picture were discussed. Here the intermultiplet-transition picture should be first excluded in these Ce$^{3+}$-based materials since the splitting between the ground multiplet $^2F_{5/2}$ and the next excited multiplet $^2F_{7/2}$ is about 250 meV~\cite{Eldeeb2020}. The proposed chemical-impurity origin of the extra INS peak in powder-sample KCeS$_2$~\cite{Bastie2020,Kulbakov2021} also cannot be used to explain the extra peak in \ccs. As discussed in Ref.~\cite{Kulbakov2021}, it requires $\sim$15.7\% Ce$_2$O$_2$S impurity to give rise to the observed intensity of the extra peak. Similarly, if the chemical-impurity scenario is applied to our case here, the Ce$_2$O$_2$Se impurity in the sample would be $\sim$15\%. However, different from the powder KCeS$_2$ sample in Ref.~\cite{Kulbakov2021}, our \ccs~sample consists of high-quality single crystals, in which such a large amount of any impurity is impossible to exist.
The CEF-phonon coupling picture seems to be also impossible in~\ccs, because all the phonon intensities from the sample are below $\sim$30 meV and there is no signature of the phonon branches around the energy scale of 50 meV.

\begin{table}[t]
\caption{Crystallographic data for CsCeSe$_2$ determined by neutron powder diffraction.}
\begin{ruledtabular}
\begin{tabular}{c c c c }
 Lattice    & $R\bar{3}m$ (No. 166)     &  $a$ = 4.40410(18) \AA   &   $c$ = 24.8546(11) \AA  \\ \hline
 atom       & $x$         & $y$          & $z$                         \\
 Cs        & 0.66667     & 0.33333      & 0.33333                      \\
 Ce        & 0.00000     & 1.00000      & 0.50000                      \\
 Se       & 0.33333     & 0.66667      & 0.43566(9)                    \\
\end{tabular}
\end{ruledtabular}
\label{Structure}
\end{table}

The calculated composition for the $^2F_{5/2}$ multiplet is summarized in Table \ref{Table:Th1}. The first and the third doublets consist of linear combinations of $|\pm1/2\rangle$ and $|\mp5/2\rangle$ states, while the doublet at 34.7 meV is a combination of $|\pm3/2\rangle$ states. The Kramers ground-state doublets of the rare-earth-based compounds, while typically transforming precisely according to the symmetry rules of an $S = 1/2$ spin, thus commonly referred to as ``effective spin-1/2'' states, may instead correspond to the higher-symmetry dipole-octupole (DO) doublets that transform differently under the point group. The higher symmetry of these doublets may result in significant differences of the effective low-energy ``spin-spin'' Hamiltonian from that for the more conventional ``spin-1/2'' doublets. Such DO ground-state doublets were recently predicted and later observed in several Ce-based pyrochlore materials~\cite{li2017symmetry, Gaudet2019, sibille2020quantum}.
Similar DO doublets were also discussed in the context of TL materials~\cite{li2016hidden} and it was shown that the DO doublet should be built up from $|\pm{}n3/2\rangle$ states ($n$ is an odd integer), while other Kramers doublets behave as standard dipoles. Our calculations [Table~\ref{Table:Th1}] clearly indicate that the ground-state doublet is a linear combination of $|\pm1/2\rangle$ and $|\mp5/2\rangle$ and thus we can safely use $S = 1/2$ approximation to describe magnetic behavior in \ccs\ at low temperatures, $T < 10$~K.

\subsection{Magnetic order from powder and single-crystal diffraction} \label{sec:diff}

\begin{figure*}[tb]
\center{\includegraphics[width=1\linewidth]{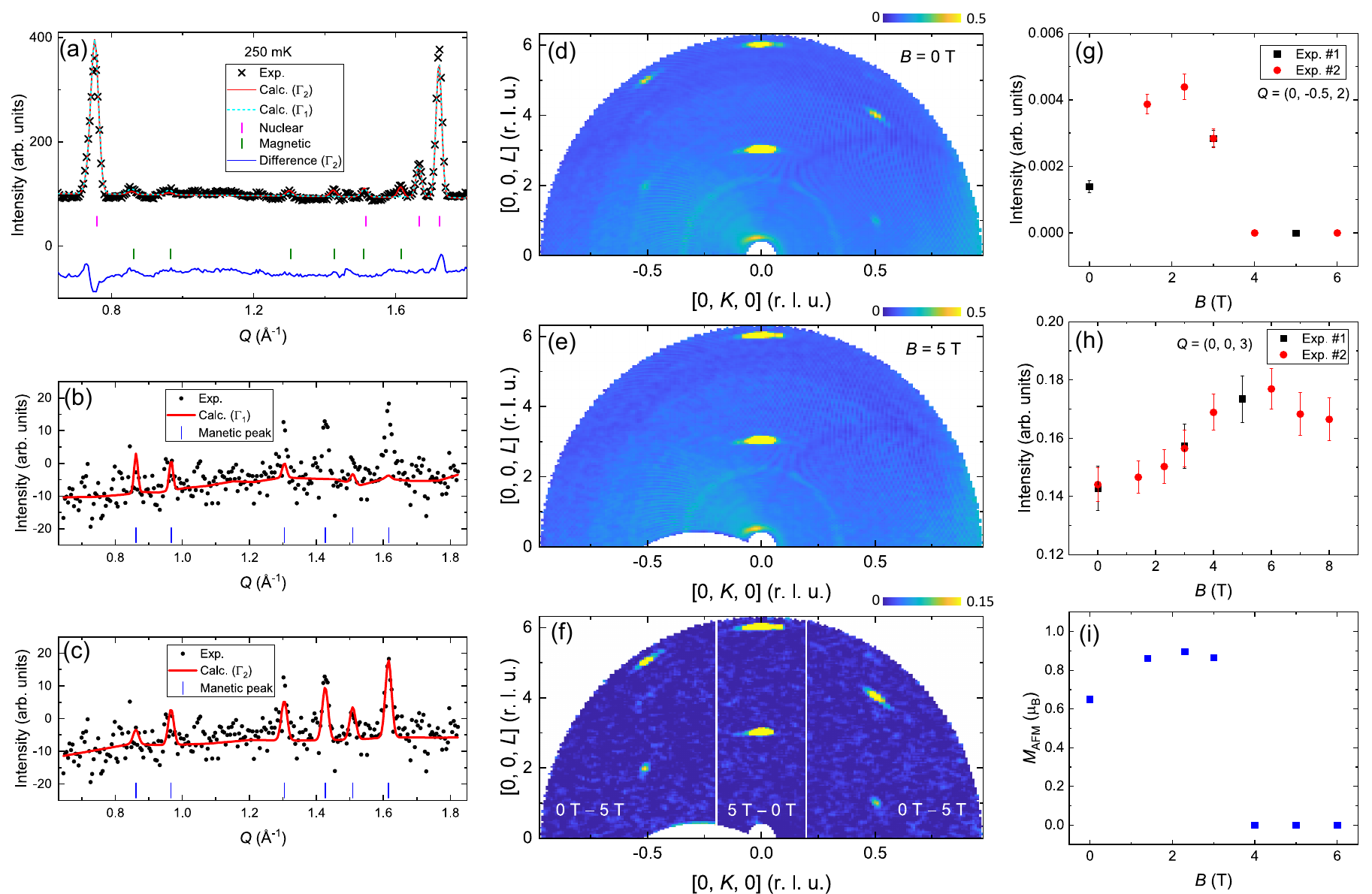}}
  \caption{~Neutron diffraction results of \ccs.
  (a)--(c)~Powder diffraction results collected at HB-2A. In panel (a), the black crosses are experimental data points measured at 250 mK, the red curves are Rietveld refinements with stripe-$yz$ ($\Gamma_2 $) magnetic order, the blue curves are the differences of the data and refinement. The magenta and olive tick marks indicate the nuclear and magnetic peak positions. The cyan dashed line represents the Rietveld refinement from the irreducible representation $\Gamma_1 $, i.e., the stripe-$x$ order. Panels (b) and (c) show the difference between 250 mK and 550 mK and refinements with stripe-$x$ ($\Gamma_1 $) and stripe-$yz$ ($\Gamma_2 $) magnetic states.
  (d)--(i)~Single-crystal diffraction data measured on CNCS at 70 mK in a series of magnetic fields.
  In panel (f), the data are obtained by subtracting 0-T dataset from 5~T one, $I(\mathbf{Q}) = I(\mathbf{Q})^{B=5~T} - I(\mathbf{Q})^{B=0~T}$ at the central part of the plot and by subtraction of 5-T dataset from 0-T for the left and right parts, $I(\mathbf{Q}) = I(\mathbf{Q})^{B=0~T} - I(\mathbf{Q})^{B=5~T}$, as indicated in the plot.
  Diffraction pattern at zero field shows AFM Bragg peaks at $\mathbf{Q}$ = (0, 1/2, 1$\pm 3n$) and $\mathbf{Q}$ = (0, $-$1/2, 2$\pm 3n$) ($n$ = integer)  that correspond to the stripe-$yz$ magnetic ground state. Strong positive intensities at $\mathbf{Q}$ = (0, 0, $3n$) indicate the FM intensity of Bragg peaks in the field-polarized (FP) state at 5 T.
    }
  \label{Diffraction}
\end{figure*}

Having determined the single-ion ground state we turn to the description of magnetic order in \ccs\ based on the results of neutron diffraction. Figure~\ref{Fig:intro}(d) presents the powder neutron diffraction result collected in the paramagnetic state. The refined crystallographic data are in agreement with previous reports~\cite{CCDC1952065, Xing20191} and are presented in Table~\ref{Structure}. Figure~\ref{Diffraction} summarizes the powder and single-crystal neutron diffraction results, which are used to characterize the magnetic ground state. Figures~\ref{Diffraction}(a)--\ref{Diffraction}(c) show the powder neutron diffraction results from HB2A in a narrow $|\bf{Q}|$ range, where the magnetic signal was observed. Below $T_{\rm N} = 0.35 \, \mathrm{K}$, we found several weak but clear magnetic Bragg peaks that can be seen most clearly in the difference profile, $I(T = 250$~mK$) - I(T = 550$~mK), see Figs.~\ref{Diffraction}(b) and~\ref{Diffraction}(c). The magnetic peaks can be indexed with a propagation wavevector $\mathbf{k} = (0~1/2~1)$. Symmetry analysis was performed using $\textsc{Mag2Pol}$~\cite{qureshi2019mag2pol} and $\textsc{SARAh}$~\cite{wills2000new} softwares and it yielded two irreducible representations, $\Gamma_1 $ and $\Gamma_2 $ (see details in Table~\ref{Irreducible}). These irreducible representations correspond to the stripe-$x$ (moments along the $a$ axis) and stripe-$yz$ orders (moments have in-plane component perpendicular to $a$ axis and an out-of-plane component), respectively and are consistent with theoretical predictions for a two-dimensional triangular lattice AFM~\cite{zhu2018topography, maksimov2019anisotropic}.

In order to distinguish $\Gamma_1$ and $\Gamma_2$ solutions we refined the raw diffraction data collected $T = 250$ mK that contain both nuclear and magnetic contributions [Fig.~\ref{Diffraction}(a)] as well as the difference profile, where the nuclear contribution is subtracted out [Figs.~\ref{Diffraction}(b) and~\ref{Diffraction}(c)]. Both approaches suggest that $\Gamma_2$ model provides a better description of the observed data. This can been seen most clearly for the refinement of the difference profile, [Figs.~\ref{Diffraction}(b) and~\ref{Diffraction}(c)], which clearly indicate that $\Gamma_2$ solution ($\chi^2 = 1.27, R_{\mathrm{F, Mag}} = 19.2$) provides much more accurate description of the experimental data as compared to $\Gamma_1$ ($\chi^2 = 2.39, R_{\mathrm{F, Mag}} = 58.3$) demonstrating significant difference in the fit quality factors.
The difference of the two solutions can also be seen on the qualitative level of comparison, e.g., magnetic peaks (0, 1/2, 1) [0.86~\AA$^{-1}$] and (0, $-$1/2, 2) [0.96~\AA$^{-1}$] have stronger intensities than the peaks (1, $-$1/2, 0) [1.41~\AA$^{-1}$]  and (1, $-$1/2, 3) [1.62~\AA$^{-1}$], in a clear contradiction with the experimental data.
Thus, we conclude that the magnetic structure of \ccs\ can be well described using stripe-$yz$ model ($\Gamma_2$ irreducible representation). Our refinement yields that the Ce$^{3+}$ magnetic moments predominantly lie within the $ab$ plane with $m_{ab}$ = 0.67(9)~\mb, having a statistically negligible out-of-plane component $m_{c}$ = 0.10(14)~\mb.

\begin{table}[tb]
\caption{Irreducible representations ($\Gamma$), the associated basis vectors ($\psi$), corresponding magnetic space groups for the $R\overline{3}m$ paramagnetic space group, with propagation vector $\mathbf{k} = (0,~1/2,~1)$. The basis vectors denote the relative moment components for the Ce site, along the three crystallographic directions in the trigonal unit cell.}
\begin{ruledtabular}
\begin{tabular}{c c c c}
    \multicolumn{1}{c}{$\Gamma$} & \multicolumn{1}{c}{$\psi$} & Components of $\psi$ & \multicolumn{1}{c}{Magnetic space group} \\	\hline
    \multicolumn{1}{c}{$\Gamma_1 $}  &                        &                      & $P_C2/m$     \\
									 & $\psi_1$               & $(4,0,0)$            &              \\
    \multicolumn{1}{c}{$\Gamma_2 $}  &                        &                      & $P_A2_1/c$   \\
									 & $\psi_2$               & $(2,4,0)$            &              \\
									 & $\psi_3$               & $(0,0,4)$            &              \\

\end{tabular}
\end{ruledtabular}
\label{Irreducible}
\end{table}

The stripe-$yz$ magnetic ground state is further confirmed by our single-crystal diffraction results. As shown in the 2D zero-energy slices in the (0 $K~L$) plane in Figs.~\ref{Diffraction}(d) and~\ref{Diffraction}(f), except for the nuclear Bragg peaks (003) and (006), we can identify additional peaks at $\mathbf{Q}$ = (0, 1/2, 1$\pm 3n$) and $\mathbf{Q}$ = (0, $-$1/2, 2$\pm 3n$) ($n$ = integer). Positions of these peaks further confirm the propagation vector $\mathbf{k} = (0,~1/2,~1)$ refined from the powder diffraction data.
The refinement of the single-crystal diffraction results with stripe-$yz$ model at zero field and 70~mK gives $m_{ab}$ = 0.65(5)~\mb, $m_{c}$ = 0.08(5)~\mb, which is similar to the values from the refinement of the powder diffraction. We use the total $m_{\rm Ce}$ = 0.65(5)~\mb from single-crystal diffraction refinement to represent the ordered moment of Ce in this paper. We note that our analysis of the zero-field magnetic structure yields a similar structure with powder neutron diffraction results reported on isostructural KCeS$_2$~\cite{Kulbakov2021}. The main difference is the stacking sequence of the triangular layers: in KCeS$_2$ the propagation wavevector was found to be (0, 1/2, 1/2)~\footnote{Authors of~\cite{Kulbakov2021} report the propagation wavevector as $\bf k_{\rm m}$ = (-1/2, 0, 1/2) which is another arm of the same $\bf k$-vector star.}, while we unambiguously determine $\bf k$ = (0, 1/2, 1) for \ccs.

\begin{figure*}[tb]
\center{\includegraphics[width=1\linewidth]{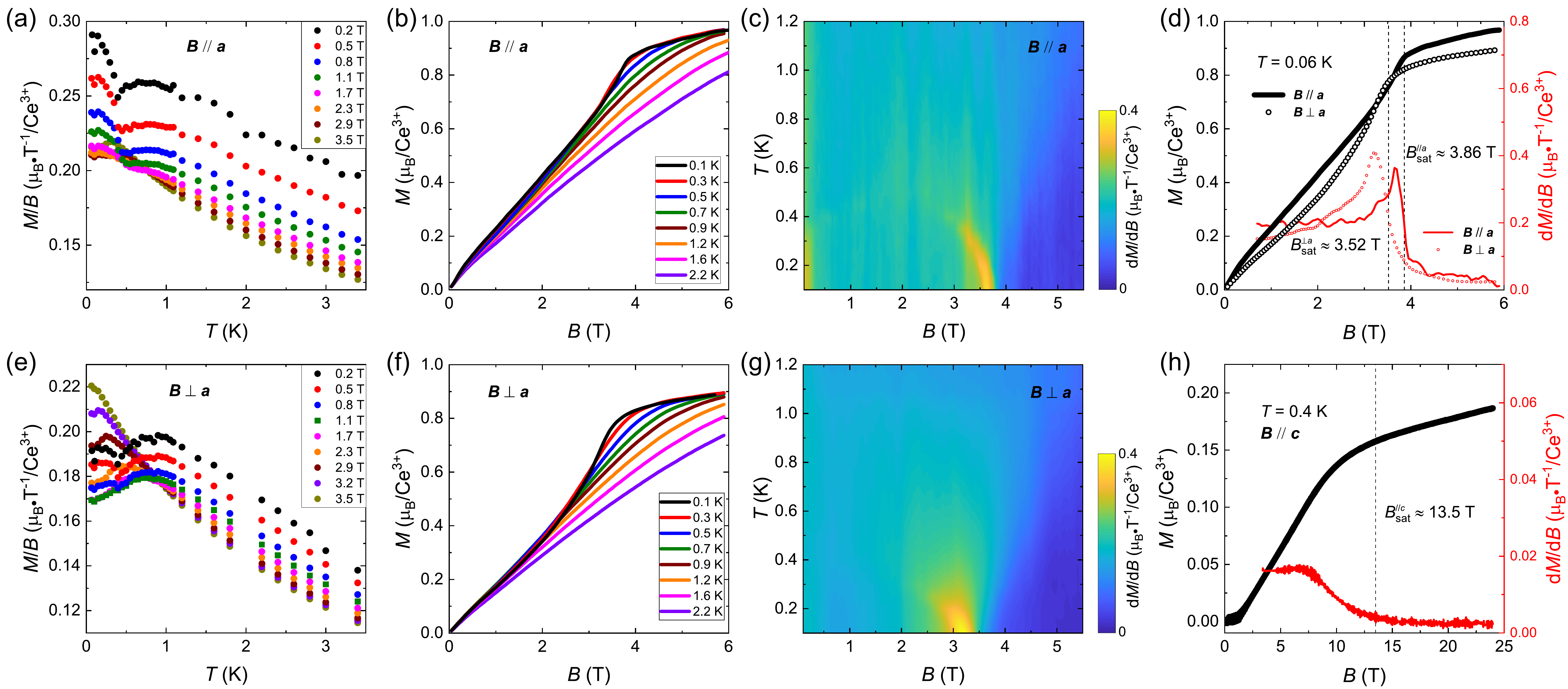}}
  \caption{~Low-temperature magnetization measurements of \ccs. Panels (a)--(c) and (e)--(g) demonstrate data collected with magnetic field applied along [110] and [100] directions.
  Panels (a) and (e) present the temperature dependence of static spin susceptibility, $M(T)/B$, measured at several fields up to saturation.
  Panels (b) and (f) show the field dependence of magnetization, $M(B)$.
  Panels (c) and (g) are the distribution of ~$dM/dB$ as a function of temperature and magnetic field.
  The weak intensity at low fields at $B<0.2$~T in panel (c) is an external artefact.
  Panels (d) and (h) present the anisotropy of magnetization in \ccs. Panel (d) shows ~isothermal magnetization as a function of the magnetic field along and perpendicular to the $a$ axis in $ab$ plane. The red curves are the first derivatives of $M(B)$. The dashed lines mark the saturation magnetic field.
  (h)~Isothermal magnetization curve measured under pulsed field with magnetic field along the $c$ axis.
  }
  \label{Fig:magnetization}
\end{figure*}

After applying magnetic fields, the intensity of these magnetic peaks first increases and reaches a maximum around $B$ = 2~T, above which these peaks are suppressed by the increasing magnetic field and disappear above $B$ = 4~T. Figure~\ref{Diffraction}(g) summarizes the evolution of the integrated intensity of the peak at $\mathbf{Q}$ = (0, $-$1/2, 2) in the magnetic fields. The intensity is suppressed to zero at $B \ge$ 4 T. At the same time, the field-induced FM intensity appears on the top of the nuclear peaks. Figure~\ref{Diffraction}(h) shows the integrated intensity of the peak at $\mathbf{Q}$  = (0, 0, 3) as a function of magnetic field, in which the intensity smoothly increases with increasing field and reaches a saturation above the saturation field ($B_{\mathrm{sat}}^{\| a}\approx{}3.86$~T is deduced from thermodynamic results in Sec.~\ref{sec:thermodynamics}). The 2D zero-energy slice at $B$ = 5~T in Fig.~\ref{Diffraction}(e) only shows the strong intensities at nuclear peaks (0, 0, 3) and (0, 0, 6), which is the result of a field-induced saturation.

To present the aforementioned AFM and FM peaks more vividly, we make subtractions between the data at $B$ = 0 T and 5 T. The resulting pattern of the zero-energy slice in the (0 $K~L$) plane is shown in Fig.~\ref{Diffraction}(f) and is divided into three parts by the two vertical white lines. The left and right parts represent the results of the subtraction of the 5-T data from the 0-T data and give clear intensities at $\mathbf{Q}$  = (0, 1/2, 1$\pm 3n$) and $\mathbf{Q}$  = (0, $-$1/2, 2$\pm 3n$) ($n$ = integer) which are the AFM Bragg peaks of the stripe-$yz$ magnetic ground state at zero field. Inversely, the middle part of Fig.~\ref{Diffraction}(f) presents the subtraction of the 0-T data from the 5-T data and shows peaks at $\mathbf{Q}$ = (0, 0, $3n$). These extra scattering intensities at the nuclear peak position represent the pure intensities of the magnetic Bragg peaks of the field-polarized~(FP) state at 5 T. The refined values of the AFM moment at different magnetic fields are summarized in Fig.~\ref{Diffraction}(i).
It shows a moderate increase to $\approx$0.86$\mu_{\rm B}$ at 2.3 T before the sharp drop to zero at $B_{\rm c}$.
Unfortunately, because of limited number of collected peaks our data do not allow for the full refinement of the magnetic structure at all fields, but we speculate that the increase of the ordered moment is the most plausible scenario consistent with enhanced transition temperature and strong peak at the phase transition observed in the specific heat measurements for intermediate fields which are presented below, see Sec.~\ref{sec:thermodynamics}.

\subsection{Thermodynamic measurements} \label{sec:thermodynamics}

\begin{figure*}[tb]
\center{\includegraphics[width=0.8\linewidth]{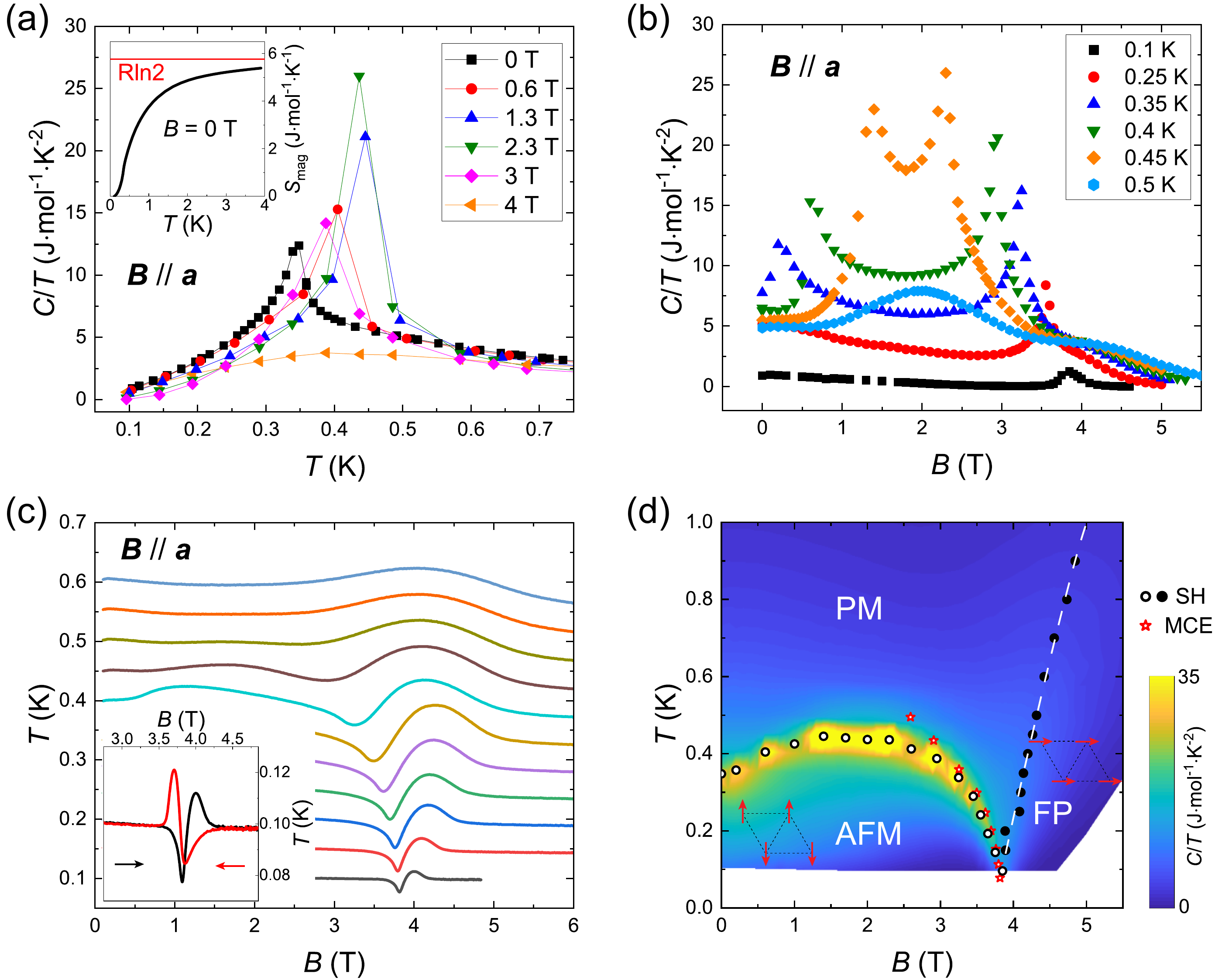}}
  \caption{~Thermodynamic characterizations and phase diagram of \ccs\ with field applied along [100] axis.
    (a)~The normalized specific heat $(C/T)$ as a function of temperature at different magnetic fields with $\textbf{B}$ $\parallel$ $\textbf{a}$. The inset shows the calculated magnetic entropy as a function of temperature at $B$ = 0 T.
    (b)~The specific heat as a function of magnetic field at different temperatures with $\textbf{B}$ $\parallel$ $\textbf{a}$.
    (c)~The MCE of \ccs\ measured with ramping up the magnetic fields. The inset is a comparison between the MCE with ramping up (black) and ramping down (red) the magnetic fields at $T$ = 0.1~K.
    (d)~A magnetic $B-T$ phase diagram, deduced from the specific heat and MCE results. The open and solid circles are data points extracted from peak positions of the specific heat (SH) data, which indicate the transitions from the paramagnetic (PM) state to the AFM state, and the crossover from the PM to the FP state. The open stars are the data points obtained from the MCE in panel (c). The white dashed line indicates the crossover between the PM and FP phases. The arrows on triangles show the sketches of the corresponding ordered states.
  }
  \label{Fig:specific_heat}
\end{figure*}

To characterize the field-induced phase diagram of \ccs\ we conduct a series of low-temperature thermodynamic measurements, including magnetization, specific heat, and MCE. Figures~\ref{Fig:magnetization}~(a) and~\ref{Fig:magnetization}(e) show the temperature dependence of the static spin susceptibility $M/B$ collected for the magnetic field applied parallel and perpendicular to the $a$ axis of a crystal, respectively. The magnetization exhibits qualitatively similar behavior in both directions. The curves collected at low fields, $B < 1$ T, show a broad maximum around 1 K that corresponds to the development of short-range correlations and a small kink around 0.4 K as a result of the establishment of the long-range AFM order. The application of magnetic field shifts the broad maximum towards lower temperatures. The second anomaly becomes less pronounced at the intermediate field range, which prevents us from accurate determination of the phase boundary $T_{\rm N}(B)$.

The temperature dependence of magnetization can also be analyzed using Curie-Weiss law, which yielded $\Theta_{\rm CW}^{ab} = -2.41$~K and $\Theta_{\rm CW}^{c} = -2.69$~K~\cite{Xing20191}. Using the equation for TLAF AFM, $\Theta_{\rm CW} = -\frac{3}{2}J/k_{\rm B}$~\cite{Ranjith2019naybse} we found that the energy scale of exchange interaction should be of the order of $\approx$0.14 meV, in qualitative agreement with the INS results~\cite{Xie2023quantum}. However, we note such analysis of susceptibility usually provides rather rough estimation of the exchange Hamiltonian especially for systems like \ccs, that exhibit both, single-ion and exchange anisotropy.

Field dependence of magnetization measured at multiple temperatures are shown in Figs.~\ref{Fig:magnetization}(b) and~\ref{Fig:magnetization}(f). The curves collected below $T_{\rm N}$ demonstrate a nonmonotonic behavior with a clear kink just below the saturation fields.
At $T > T_{\rm N}$ the magnetization shows regular Brillouin-like behavior as expected for the paramagnetic (PM) phase.
Figures~\ref{Fig:magnetization} (c) and~\ref{Fig:magnetization}(g) present the color maps of $dM/dB$ for both in-plane field directions ($\textbf{B}$~$\|~\textbf{a}$ and $\bot~\textbf{a}$), where the field-induced transition between stripe-type AFM and FP state can be seen at low temperatures ($T < 0.4$~K).
The magnetization measured along two principal in-plane directions shows considerable anisotropy, which can be seen in the $M(B)$ curves taken at $T$ = 0.06 K in Fig.~\ref{Fig:magnetization}(d). The saturation fields differ substantially with $B_{\mathrm{sat}}^{\| a}\approx3.86$~T and $B_{\mathrm{sat}}^{\bot a}\approx3.52$~T. Such anisotropy of saturation field is not expected for the TLAF with isotropic or XXZ exchange interactions and proves the presence of the bond-dependent terms in the spin Hamiltonian. Our results indicate that the ratio of saturation fields $B_{\mathrm{sat}}^{\| a}/B_{\mathrm{sat}}^{\bot a}>1$, which is consistent with the stripe-$yz$ ground state~\cite{avers2021finger} deduced by diffraction measurements in Sec.~\ref{sec:diff}. The magnetization curves above the saturation show a minor ($<$10\%) difference of saturation moment, which we associate with the uncertainty of mass for small samples ($<$1 mg). Moreover, the linear growth of magnetization above $B_{\rm sat}$ can be explained as the van Vleck contribution, which has been observed in multiple materials of this family~\cite{Ranjith2019naybse,xie2023complete}.

Figure~\ref{Fig:magnetization}(h) displays the field dependence of the magnetization measured at $T=0.4$~K in pulsed magnetic fields up to 24~T applied parallel to the $c$ axis. The absolute values were obtained by normalizing the data to a low-field magnetization curve measured in a MPMS. Magnetization of CsCeSe$_2$ shows stronger anisotropy between the $ab$ plane and the $c$ axis. The high-field saturation moments within the $ab$ plane are about six times larger than the moment along the $c$ axis, confirming that the Ce$^{3+}$ ground-state doublet has strong easy-plane anisotropy induced by CEF as discussed in Sec.~\ref{sec:CEF}. Thus, using the magnetization data we reveal the magnetic anisotropy, the presence of bond-dependent interactions and critical fields at the low-temperature regime.

Figure~\ref{Fig:specific_heat}(a) depicts the specific heat of \ccs\ as a function of temperature for different magnetic fields with $\textbf{B}$ $\parallel$ $\textbf{a}$. At zero field a PM-AFM transition is observed at $T_{\rm N} = 0.35(2)$~K. The integrated magnetic entropy reaches $\approx 93.5~\%$ of $R \ln 2$ (the value expected for a doublet ground state) at 4~K, which confirm the pseudo-spin-1/2 nature of Ce magnetism. With increasing field, the sharp peak observed at zero field first rises and shifts towards higher temperatures having maximum in both, temperature and intensity, around 2~$\mathrm{T}$. Further field increase suppresses the peak, and eventually, only a broad anomaly can be resolved at $B \geq 4~\mathrm{T}$. We note that this behavior is consistent with the nonmonotonic behavior of the AFM moment observed by single-crystal neutron diffraction [Fig.~\ref{Diffraction}~(i)]. The magnetic field dependence of the specific heat at several temperatures is shown in Fig.~\ref{Fig:specific_heat}(b). At the lowest temperature of 0.1~K we observe a single peak at $B = 3.86(1)$~T, which perfectly coincides with the saturation field deduced from the magnetization measurements [Fig.~\ref{Fig:magnetization}(d)]. At $T \geq \, 0.35$~K we observe two peaks, which mark the transition temperatures $T_{\rm N}(B)$, and clearly indicate reentrant behavior of the phase diagram of \ccs.
Above the critical field isothermal specific heat curves, $C(B)$, demonstrate the second peak that moves to higher fields with temperature caused by field-induced increase of the spin gap. We associate this anomaly with the crossover between PM and FP states. To extract position of this anomaly we used inflection-point fit and the high-field crossover line is shown in Fig.~\ref{Fig:specific_heat}(d).

We finalize the presentation of our thermodynamic results with the MCE, which is an appropriate method for studying field-dependent phase boundaries.  Figure~\ref{Fig:specific_heat}(c) shows the quasi-adiabatic sample temperature variation, the MCE, during the magnetization process at several representative temperatures. Positive and negative reversible changes $\delta T$ in the sample temperature are observed at the critical field $B_{\mathrm{sat}} \approx{}3.8$~T, as expected in the vicinity of the transition from the AFM to the FP phase. We find that the magnitudes of positive and negative $\delta T$ retain similar values at low temperatures [see the inset of Fig.~\ref{Fig:specific_heat}(c)] and do not show an asymmetry characteristic of a first-order transition~\cite{Kohama}, suggesting a second-order-like transition. As temperature increases, the negative anomaly broadens and shifts towards lower fields. We note that such a behavior of $T_{\rm N}(B)$ has been detailed in another stripe-ordered Ce-based TLAF CeCd$_3$As$_3$~\cite{avers2021finger}.

Figure~\ref{Fig:specific_heat}(d) summarizes a magnetic phase diagram for \ccs~ with the field applied along the $a$ axis. Our thermodynamic measurements allow us to draw several conclusions regarding the field-induced evolution of the ground state. First of all, the anisotropy of in-plane critical fields strongly suggests the presence of bond-dependent terms. No additional field-induced transition can be observed within the AFM phase by any of our experimental probes, which suggests a simple field-induced canting of the collinear spin structure at zero field towards the field direction as schematically shown in Fig.~\ref{Fig:specific_heat}(d). With the increase of the magnetic field, the N\'{e}el temperature $T_{\rm N}$ first shows considerable enhancement from 0.35 K to 0.42 K around 2 T, and then it decreases continuously to zero at critical field $B_{\mathrm{c}}^{\| a}\approx3.86$ T. These observations are perfectly consistent with the field-induced evolution of the AFM Bragg peaks discussed in Sec.~\ref{sec:diff}. The MCE and the specific heat data provide strong evidences for a magnetic-field-driven quantum critical point (QCP) at $B_{\mathrm{c}}^{\| a}$.

\subsection{Magnetic moment and $g$ tensor of Ce$^{3+}$}
\label{sec::magnetic_moment}
In this paper we report three independent methods to characterize the magnetic moment of Ce$^{3+}$ in \ccs: magnetization measurements, neutron diffraction and DFT calculations. The fourth approach, in-field INS measurements, is presented in~\cite{Xie2023quantum} and yields $g_{ab} = 1.77$ ($m = 0.88~\mu_{\rm B}$). The saturation magnetization values suggest slightly lower values of $g_{ab}$, $\sim$~1.5 and $\sim$~1.7 for $\mathbf{B}\parallel{}\mathbf{a}$ and $\mathbf{B}\perp{}\mathbf{a}$ directions, respectively. However, taking into account the uncertainty of the sample mass, the results of both INS and magnetization experiments are in a reasonable agreement.

Neutron diffraction yields an ordered moment $m_{\rm Ce}$ = 0.65(5)~$\mu_{\rm B}$ at zero field, which primarily lies within the $ab$ plane. It should be noted that neutron diffraction probes the ordered component of the magnetic moment that is known to be reduced in many frustrated systems, while INS and high-field magnetization measure the full moment of Ce$^{3+}$ ground state doublet. For instance, in the isostructural KCeS$_2$ the ordered moment from neutron diffraction was found to be $0.32~\mu_{\rm B}$~\cite{Kulbakov2021}, while the full moment measured accurately by electron spin resonance (ESR) measurements is 0.82~$\mu_{\rm B}$ ($g_{ab} = 1.65$)~\cite{Bastie2020} implying that only 40\% of the full magnetic moment participates in the magnetic order. In \ccs\ the reduction of the ordered moment at zero field is more mild and 75\% of the full magnetic moment orders at zero field. Application of magnetic field enhances the transition temperature significantly and makes the associated specific-heat peak stronger, see Fig.~\ref{Fig:specific_heat}(a). This effect correlates with moderate enhancement of the ordered moment observed in the neutron diffraction data at intermediate fields [Fig.~\ref{Diffraction}(i)]. The maximum of the ordered magnetic moment in $ab$ plane (at $B\sim$~2.3 T), 0.86~$\mu_{\rm B}$ is very close to the nominal saturation moment $m = 0.88~\mu_{\rm B}$ in $ab$ plane from the INS result with $g_{ab} = 1.77$. Thus, we conclude that all experimental approaches agree with each other and INS, as the most accurate method, yields $g_{ab} = 1.77$.
The out-of-plane component of the $g$ tensor can be extracted from the high-field magnetization measurements [Fig.~\ref{Fig:magnetization}(h)], $g_c \approx\ 0.3$.

DFT calculations of CEF yield $g_{ab} = 2.3$ and $g_c = 0.5$ having the qualitative agreement with the experimental values, $g_{ab} \gg\ g_c$.
Reports on the isostructural compound KCeS$_2$ show how the $g$ tensor values are sensitive to small changes in the local geometry of the Ce octahedra, both experimentally~\cite{Bastie2020} and theoretically~\cite{Kulbakov2021}.
That is because the $g$ tensor structure reflects the mixing of the Kramers doublets and, unlike the excitation energies. It is difficult to tune the theoretical settings and requires an independent study, which is beyond the scope of this paper.

\subsection{Theoretical considerations}
\label{sec::hamiltonian}
Our thermodynamic data presented in Sec.~\ref{sec:thermodynamics} strongly suggest that \ccs\ exhibits a field-induced QCP when the magnetic field is applied along in-plane directions, $\textbf{B} \parallel{}\textbf{a}$ and $\textbf{B} \perp{}\textbf{a}$.
To get more insight into the universality class of this QPT we make use of Ginzburg-Landau type arguments and ED calculations. We start this subsection by presenting the effective spin Hamiltonian for \ccs{}, then present the determination of the critical fields in LSWT and then discuss the nature of the phase transitions deduced from ED in the case of an isolated layer. We finish by briefly commenting on the effect of finite interlayer coupling.

\subsubsection{Spin Hamiltonian for \ccs}

\begin{figure*}
\center{\includegraphics[width=1\linewidth]{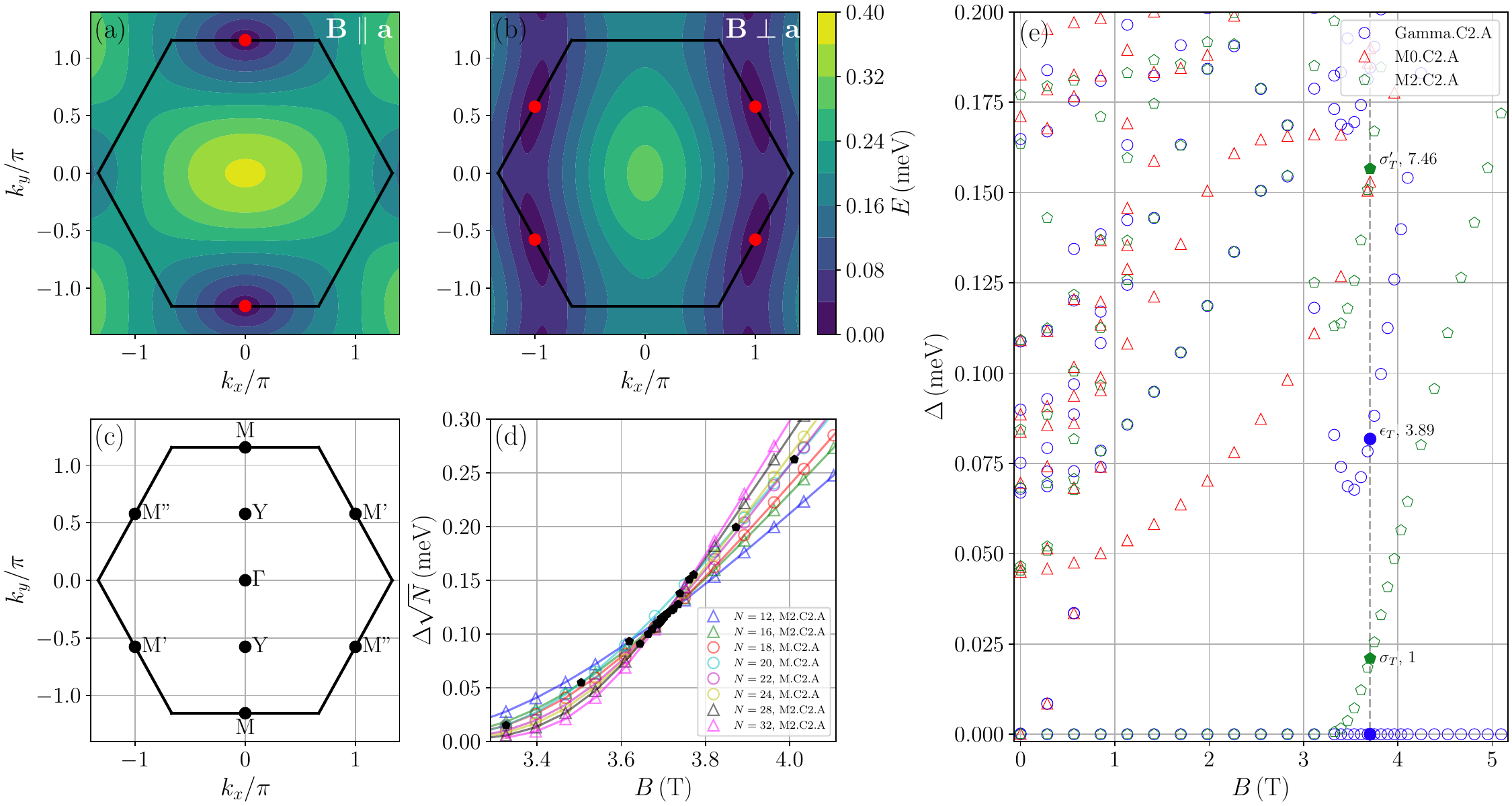}}
\caption{
  (a)(b)~LSWT dispersion just above the critical field applied (a) parallel to~${\bf a}$ ($B_{\rm c,LSWT}^{\parallel {\bf a}}=4.458 \, \mathrm{T}$) and (b) perpendicular to~${\bf a}$ ($B_{\rm c,LSWT}^{\perp {\bf a}} = 3.742 \, \mathrm{T}$). Note that LSWT overestimates $B_{\rm c}$ and the values used for these calculations are larger than experimental values and those determined by ED. The red circles show the minimum of the dispersion, which vanishes when $B=B_{\rm c}$, at M in (a) and M$'$ in (b).
  (c)~Magnetic Brillouin zone (in normalized units). When the magnetic field vanishes, the three M points, M, M$'$, and M$''$ are inequivalent, while for a nonvanishing magnetic field, the BZ is folded and M$'$ and M$''$ become equivalent.
  (d)~Finite-size gap $\Delta$ from the zero-momentum ground state to the first excited state at the M point multiplied by $\sqrt{N}$ for different system sizes $N$ in proximity to the critical field.
  (e)~Finite-size spectrum versus magnetic field of the $N=32$ sites torus, restricted to states at $\Gamma$, $\mathrm{M}\equiv \mathrm{M2}$ and $\mathrm{M'}\equiv \mathrm{M0}$. The vertical-dashed line shows the critical field as obtained in~(d), $B_{\rm c}\simeq 3.7 \, \mathrm{T}$. The filled markers show three gaps appearing at $\Gamma$ and M at $B=B_{\rm c}$, which allow us to identify the nature of the phase transition. The numerically obtained renormalized values $\Delta/\Delta_0$ are written beside each gap.}
\label{Fig:theory_all}
\end{figure*}

As demonstrated in the previous sections, \ccs{} exhibits bond-dependent interactions. We thus consider the most general Hamiltonian in a magnetic field for a spin-$1/2$ system on the triangular lattice with nearest neighbor exchange coupling~\cite{Rau2018Frustration,maksimov2019anisotropic}
\begin{equation}
\mathcal{H} = \sum_{\langle ij \rangle} \mathbf{S}_i^{\mathrm{T}} \hat{\bf J}_{ij} \mathbf{S}_j - g_{ab} \mu_{\mathrm{B}} \mathbf{B} \cdot \sum_i \mathbf{S}_i
\label{eq::def_Hamiltonian}
\end{equation}
where $g_{ab}$ is the $g$ factor in $ab$ plane, $\mu_{\mathrm{B}}$ is the Bohr magneton, $\mathbf{B} \cdot \mathbf{c} = 0$, namely the magnetic field is applied in the $ab$-plane, and $\hat{\bf J}_{ij}$ is the matrix of couplings, which depends on the bond orientation. For the bond along the $\bm{\delta}_1 = \mathbf{a}$ direction, it can be chosen as~\cite{maksimov2019anisotropic,li2015rare,Liyaodong2016}
\begin{equation}
\hat{\bf J}_{ij(\bm{\delta}_1)} = \begin{pmatrix}
J+2J_{\pm\pm} & 0 & 0 \\	
0 & J-2J_{\pm\pm} & J_{z\pm} \\
0 & J_{z\pm} & \Delta{}J \\
\end{pmatrix}
\label{eq::Jij}
\end{equation}
where $J>0$ means the antiferromagnetic Heisenberg coupling, $\Delta$ is the XXZ anisotropy, and $J_{\pm\pm}$ and $J_{z\pm}$ are the couplings of the bond-dependent interactions. For the two other bond directions, $\bm{\delta}_2$ and~$\bm{\delta}_3$, the matrix of couplings can be obtained from~\eqref{eq::Jij} with a rotation of $\pm 2\pi/3$, $\hat{\bf J}_{ij(\bm{\delta}_q)} = {\bf R}_{-2\pi (q-1)/3} \hat{\bf J}_{ij(\bm{\delta}_1)} {\bf R}_{2\pi (q-1)/3}$ where ${\bf R}_{\alpha}$ is the rotation matrix with a rotation angle $\alpha$ in $ab$ plane.
Parameters in Hamiltonian~\eqref{eq::def_Hamiltonian} for \ccs, $J = 72.5~\mu$eV, $\Delta = 0.25$, $J_{\pm \pm}/J = 0.52$, $J_{z \pm}/J = 0.41$ and $g_{ab} = 1.77$, have been determined by fitting the INS spectra using LSWT and ED in Ref.~\cite{Xie2023quantum}.

\subsubsection{LSWT calculations}

Within the LSWT approximation, one can use the analytical expressions for the magnon mode in the field-polarized phase to derive the critical field of a transition between the stripe-$yz$ and the paramagnetic phase as the field at which the gap at the corresponding ordering vector vanishes. Equivalent results should also follow from the classical energy minimization. For the two principal directions the results are given by
\begin{align}
g_{ab}\mu_B B_{\rm c,LSWT}^{\perp {\bf a}}&= (7+\Delta)J/2 +J_{\pm\pm} \nonumber \\
&+\sqrt{\big((1-\Delta)J/2+J_{\pm\pm}\big)^2+3J_{z\pm}^2},
\label{eq:LSWT_Bc}\\
g_{ab}\mu_B B_{\rm c,LSWT}^{\parallel {\bf a}}&= (7+\Delta)J/2 +2J_{\pm\pm} \nonumber \\
&+\sqrt{\big((1-\Delta)J/2+2J_{\pm\pm}\big)^2+4J_{z\pm}^2}.\nonumber
\end{align}
Using the values of the parameters above, the fields are
\begin{align}
B_{\rm c,LSWT}^{\perp {\bf a}} = & \ 3.742\ {\rm T},
\label{eq:LSWT_Bc_values}\\
B_{\rm c,LSWT}^{\parallel {\bf a}} = & \ 4.458\ {\rm T}. \nonumber
\end{align}
While these values differ from the experimental ones, these differences can be reconciled as a result of the quantum effects, which are properly accounted for by the ED calculations below. Curiously, such quantum effects appear stronger in the
$B_{\rm c,LSWT}^{\parallel {\bf a}}$ field. Most importantly, these results show explicitly the importance of the bond-dependent $J_{\pm\pm}$ and $J_{z\pm}$ terms in making the difference of the critical fields in the two principal direction non-zero.

\subsubsection{Isolated triangular layer}

We first discuss the ground state (i.e.,~$T=0$) physics of an isolated
triangular layer as a function of the strength and orientation
of the magnetic field within the triangular plane. At large magnetic fields
the spins are in the field-polarized phase, irrespective of the field orientation. At zero field on the other hand, we have three different domains of stripe antiferromagnetic order. Since the considered Hamiltonian has no continuous spin symmetry, but only discrete symmetries, we expect six different ground-state configurations in total, consisting of two spin flip related states for each of the three domain orientations.

Within LSWT, the dispersion of the magnetic excitations in the field-polarized regime can be well described. Interestingly the number of minima of the dispersion and
their location in the BZ do depend on the field orientation. As can
be seen in Figs.~\ref{Fig:theory_all}(a) and~\ref{Fig:theory_all}(b) close to the phase transitions for two particular cases,
namely the field applied parallel to $\mathbf{a}$ or perpendicular to~$\mathbf{a}$ in the plane, leads to either
a single minimum at the M point for $\mathbf{B}\parallel\mathbf{a}$, while $\mathbf{B}\perp\mathbf{a}$ leads to two degenerate minima at the M$'$ and M$''$ points respectively [see Fig.~\ref{Fig:theory_all}(c) for a sketch of the magnetic BZ].

The number of modes softening at the field-polarized to striped AFM transition
has an impact on the universality class of the QPT for the
isolated layer. Using a standard soft-mode symmetry analysis based on Ginzburg-Landau theory leads to the following scenarios: The single minimum at the $M$ point for $\mathbf{B}\parallel\mathbf{a}$ is expected to lead to a 2+1D Ising universality class. In contrast, $\mathbf{B}\perp\mathbf{a}$ with the two soft modes
leads to an effective two-component, cubic anisotropy model, like in the
2+1D Ashkin-Teller model studied, e.g.,~in Ref.~\cite{Schuler2023}. In this theory there is a free parameter, which can change the phase transition from first order, to two copies of a 2+1D Ising at a fine-tuned point, to a 2+1D XY universality class. So we see that within a Ginzburg-Landau treatment
the two field orientations lead to distinct
quantum phase transitions. As can be seen in Figs.~\ref{Fig:theory_all}(a) and~\ref{Fig:theory_all}(b), the dispersion is spatially anisotropic
even in the vicinity of the minimum with a direction dependent spin wave velocity. The asymmetry of the velocities is particularly pronounced for $\mathbf{B}\perp\mathbf{a}$. While the $\mathbf{B}\parallel\mathbf{a}$ field direction analysis is robust with respect to small deviations of the field away from the
$\mathbf{a}$ direction, the $\mathbf{B}\perp\mathbf{a}$ case is more fragile. A deviation of the
field direction will lift the degeneracy between the two M$'$ and M$''$ ordering momenta, and one of them will be selected accordingly.

In order to corroborate the nature of the transition when the field is applied parallel to~$\mathbf{a}$, we employ exact diagonalization to study the finite-size energy spectrum at the transition, as introduced in Refs.~\cite{Schuler2016,Whitsitt2017}.
We begin by verifying the scaling behavior at the transition by rescaling the finite-size energy spectrum obtained with ED by the linear system size, $\sqrt{N}$, for tori of sizes $N=12$ up to $N=32$.
Figure~\ref{Fig:theory_all}(d) shows the rescaled gap from the ground state to the first excited state at M versus magnetic field. These results strongly support the presence of a continuous transition where the spectrum collapses as $1/\sqrt{N}$ at the critical point, while the gap vanishes exponentially in the thermodynamic limit below this critical field. To accurately locate the critical field, we identify the crossing point between each pair of system sizes. Averaging over the crossing fields we obtain $B_c \simeq 3.7 \, \mathrm{T}$, in very good agreement with the experimental value.
We can now analyze the finite-size energy spectrum at the transition. Figure~\ref{Fig:theory_all}(e) shows the spectrum of the $N=32$ sites cluster, restricted to states at the $\Gamma$ and M points. At the critical field, we compute the rescaled gaps $\Delta/\Delta_0$ for the three first excited states at the $\Gamma$ and M points with $\Delta_0$ as the gap from the ground state to the first excited state at M. Given the small number of cluster sizes accessible with ED, and unlike in Ref.~\cite{Schuler2016}, we do not perform any finite-size extrapolation, but rather fix $\Delta_0$ to its $N=32$ value. The second rescaled gap $\Delta/\Delta_0 \simeq 3.89$ is in good agreement with the corresponding value for the $\epsilon_T$ field of the 2+1D Ising universality class~\cite{Schuler2016}, given the spatial anisotropy of the velocity. The agreement between the numerical value at $N=32$ ($\Delta/\Delta_0 = 7.46$) and the theoretical value for the $\sigma_T'$ field is less striking. However, this can be explained by finite-size effects as this gap has a stronger dependence on $1/N$ than the first two gaps~\cite{Schuler2016}. We note that the particular field dependence of the quantum critical points could also be explored in future experiments, where the spectral functions in the vicinity of the ordering wavevector(s) would exhibit broad response with the absence of quasiparticle excitations.

Next, we explore the finite-temperature phase transitions in a single layer,
when the magnetic field is finite, and a single domain of the stripe AFM phase is selected. In that case we expect the system to exhibit a simple 2D Ising phase transition between the stripe and the paramagnetic phase. At zero field, or when the field is perpendicular to $\mathbf{a}$, the ground-state degeneracy is higher (six- or fourfold), and the finite-temperature transitions can possibly belong to several different universality classes and scenarios beyond
the 2D Ising case, see~Ref.~\cite{Smerald2016}. We leave a detailed study of the 2D thermal phase diagram for future work.

\subsubsection{Coupled layers, three-dimensional situation}
In the experimental compound, the magnetic triangular layers have an ABC stacking. Assuming a stripe order in each layer, then antiferromagnetic closest neighbor interactions between the layers prefer a definite ordering pattern from layer to layer, despite some inherent frustration. So the ground-state degeneracy remains the same in the three-dimensional situation compared to the isolated layer considered above. However, the increased spatial dimensionality alters the universality classes.  For the field in the $\mathbf{a}$ direction and zero temperature we expect a phase transition in the $3+1$D Ising universality, which is the same as mean field theory with Gaussian fluctuations. The thermal transitions are of the 3D Ising universality class. Similar considerations apply for the zero-temperature case with the field perpendicular to $\mathbf{a}$ and the thermal transition in the zero-field case.

\section{Discussion and Conclusions}\label{sec:disc}

In this paper, we report a comprehensive characterization of the magnetic behavior in the ideal TLAF \ccs. The high-energy INS data, low-temperature specific heat, and first principle calculations indicate that the CEF splitting is large compared to the exchange interaction energy and the low-temperature properties are dominated by the ground-state doublet that can be effectively viewed as a pseudo-spin-$1/2$ degree of freedom.

Single-crystal and powder neutron diffraction data demonstrate that the ground state is a stripe-$yz$ AFM state, rather than the famous 120$^{\circ}$ state, expected for a distortion-free TLAF with isotropic nearest-eighbor couplings.
Application of magnetic field first increases the ordering temperature from 0.35 K at zero field to 0.45 K around 2 T and eventually suppresses the stripe order in favor of the FP ferromagnetic state. Notably, anisotropy of the saturation field deduced from magnetization measurements, together with stripe-type magnetic order, suggest the presence of the anisotropic BD terms in the spin Hamiltonian.

The BD terms have profound consequences on the physical behavior of \ccs. First, in the absence of next-nearest-neighbor interaction, it is responsible for the stabilization of the stripe ground state.
Second, the BD interactions make the universality class of the QCP for the field applied along and perpendicular to the $a$ axis nonequivalent. The phase transition between AFM and FP phases can be seen as a condensation of the magnons at two (four) M points of the BZ for field applied along (perpendicular) to the $a$ axis.
In the first case, the system exhibits a transition in the 2+1D Ising universality class at zero temperature that changes to a simple 2D Ising phase transition at finite temperatures.
For $\textbf{B}\perp{}\textbf{a}$ the situation is more complicated because the ground-state degeneracy is higher (six- or fourfold). Our soft-mode symmetry analysis shows that this can lead to various scenarios of QPT, including two copies of 2+1D Ising or 2+1D XY universality classes, depending on a free parameter of the model.
Moreover, the BD terms lead to nontrivial modification of the excitation spectrum and strong interaction between a single-magnon branch and two-magnon continuum that are addressed in details in~\cite{Xie2023quantum}.
To conclude, our paper provides a comprehensive study of the low-temperature magnetic properties of \ccs, and suggests a path towards a more complete understanding of frustrated quantum magnetism.

\section*{Acknowledgements}
We thank Dr. Hongtao Liu (Instrument Analysis \&\ Research Center, Sun Yat-sen University) for the assistance with LA-ICP-TOF measurements and data evaluations. We thank Dr. Jong Keum for the help with the x-ray Laue and x-ray diffraction measurements, Dr. Feng Ye for assistance with single-crystal x-ray diffraction measurements, Dr. J\"org Sichelschmidt for the attempts of the ESR measurements and for the helpful discussion.
Work at Sun Yat-sen University was supported by the National Natural Science Foundation of China (Grant No. 12304187), the open research fund of Songshan Lake Materials Laboratory (Grant No. 2023SLABFN30), the Guangzhou Basic and Applied Basic Research Funds (Grant No. 2024A04J4024), and the Fundamental Research Funds for the Central Universities, Sun Yat-sen University (Grant No. 23qnpy57).  Work at Oak Ridge National Laboratory (ORNL) was supported by the U.S. Department of Energy (DOE), Office of Science, Basic Energy Sciences, Materials Science and Engineering Division. This research used resources at the Spallation Neutron Source and the High Flux Isotope Reactor, DOE Office of Science User Facilities operated by the Oak Ridge National Laboratory.
x-ray Laue and XRD measurements were conducted at the Center for Nanophase Materials Sciences (CNMS) (CNMS2019-R18) at ORNL, which is a DOE Office of Science User Facility. The pulsed magnetic field magnetometry measurements at National High Magnetic Field Laboratory are supported by the U.S. DOE, Office of Science, via BES program ``Science of 100 Tesla".
The work of A.L.C. on the analytical theory was supported by the U.S. Department of Energy, Office of Science, Basic Energy Sciences under Award No. DE-SC0021221.

\appendix
\section{Single-Crystal x-ray Diffraction}

\begin{figure}[t]
\center{\includegraphics[width=\columnwidth]{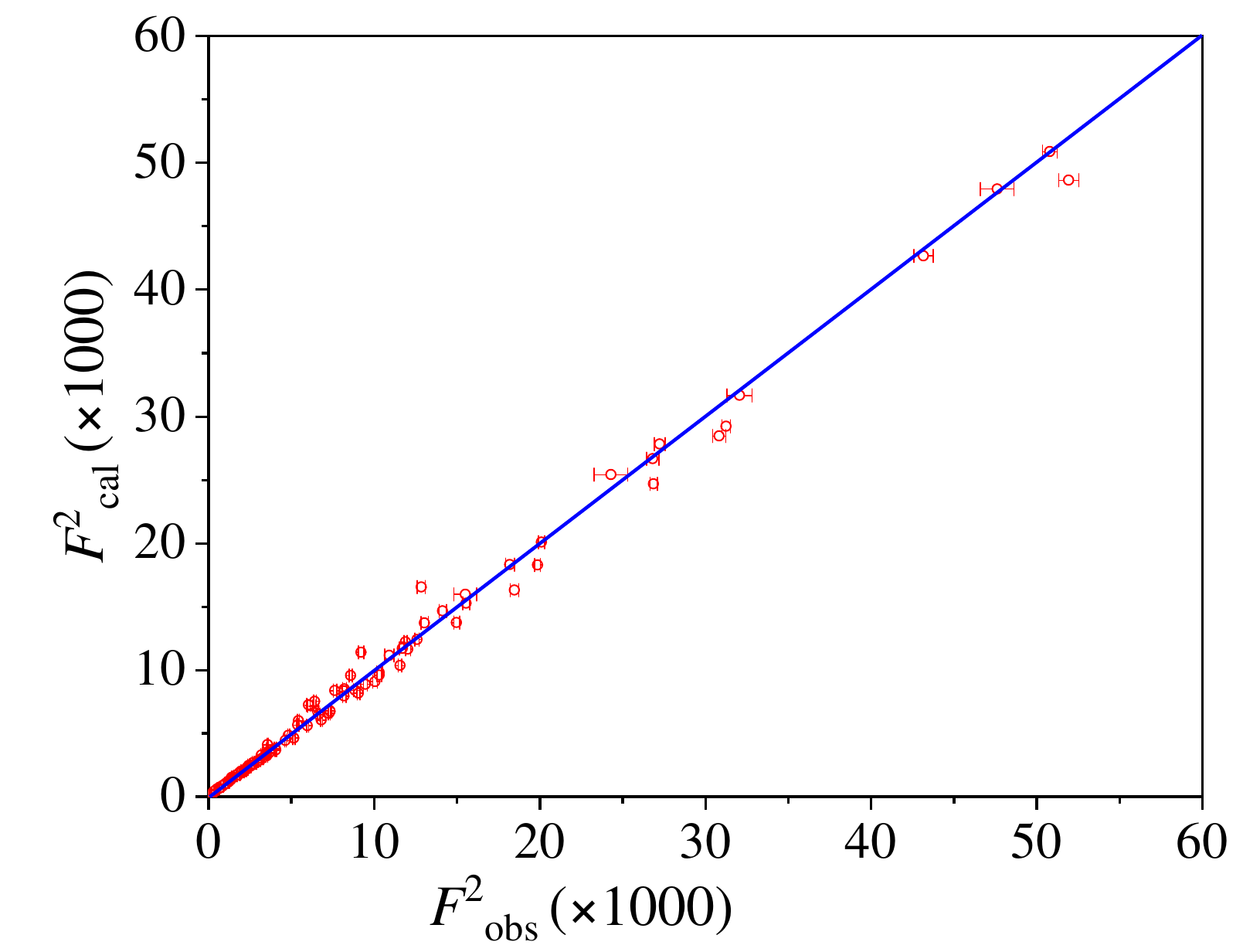}}
\caption{Rietveld refinement of single-crystal x-ray diffraction data. $F^2_{\mathrm{obs}}$ and $F^2_{\mathrm{cal}}$ denote respectively the observed and calculated structure factors.}
\label{SCXRD}
\end{figure}

The lattice structure of our CsCeSe$_2$ samples have been characterized by single-crystal x-ray diffraction on several batches of single crystals using a Bruker Quest D8 single-crystal x-ray diffractometer. The details about the refinements and crystallographic data can be found in Ref.~\cite{Xing20191}. The crystal structure has been deposited in the Cambridge Crystallographic Data Centre (CCDC)~\cite{CCDC1952065}. The single-crystal x-ray diffraction indicates the high quality of our single crystals. Here we append a result of a Rietveld refinement of single-crystal x-ray diffraction data in Fig.~\ref{SCXRD}.

\section{Chemical analyses: LA-ICP-TOF and EDS Measurements}

\begin{table}[t]
\caption{The experiment conditions for LA-ICP-TOF analysis.}
\begin{ruledtabular}
\begin{tabular}{c c}
 Conditions for laser ablation    &   \\ \hline
 He gas flow (L/min)       & 0.7      \\
 Laser fluence (J/cm$^2$)  & 1.33   \\

 Spot size ($\mu$m)        & 30     \\
 Scan speed ($\mu$m/s)        & 30     \\
 Repetition frequency (Hz)       & 10    \\
 Ablation mode       & line-scan   \\ \hline
 Conditions for ICP-TOF   &     \\ \hline
 Nebulizer gas flow (L/min)      &       0.6                   \\
 Auxiliary gas flow (L/min) &       0.8                   \\

 Plasma gas flow (L/min)      & 14.0                      \\
 RF Power (W)      & 1550                      \\
 Sample depth (mm)      & 5                    \\
 Integration time (ms)      & 500                    \\
 Data acquisition (s)     & 30  \\
\end{tabular}
\end{ruledtabular}
\label{ICP}
\end{table}

\begin{figure*}[tbh]
\center{\includegraphics[width=1\linewidth]{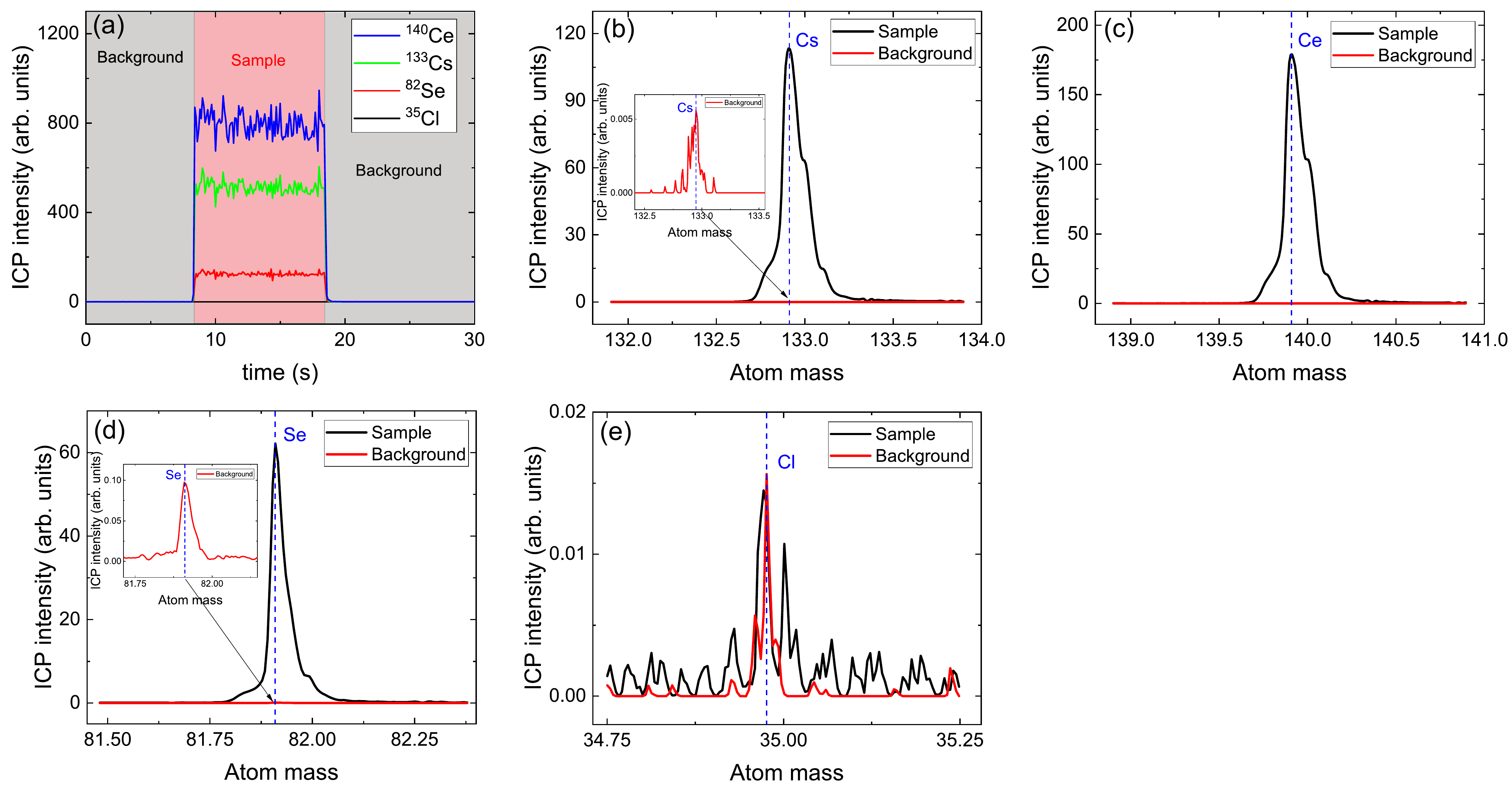}}
\caption{Time-resolved ICP results (a) and the mass spectra [(b)-(d)] of the CsCeSe$_2$ single crystals.}
\label{Fig-ICP}
\end{figure*}

The femtosecond laser-ablation hyphenated with inductively coupled plasma time-of-flight mass spectrometry (LA-ICP-TOF) was used to figure out whether the CsCeSe$_2$ crystals were contaminated by Cl from the flux during the growth process. For LA-ICP-TOF measurements, a Yb:YAG femtosecond laser ablation system (NWR-femto, Elemental Scientific Instruments, USA) working at 257~nm was coupled to an ICP-TOF mass spectrometer (icpTOF R, TOFWERK, Switzerland). The laser ablation system was working in line-scan mode and the aerosols were transported via a helium flow to the ICP-TOF for element analysis. Time-resolved signal as well as mass spectra were recorded simultaneously. The internal trigger was started prior to laser ablation so that the background signals and signals from samples could be collected and compared simultaneously. The experiment condition was optimized as listed in Table~\ref{ICP}. We present the ICP results in Fig.~\ref{Fig-ICP}. The time resolved intensities of $^{35}$Cl, $^{133}$Cs, $^{140}$Ce, $^{82}$Se were shown in Fig.~\ref{Fig-ICP}(a), in which the grey regions represent the signals without the sample, i.e., background signals, and the red area represents the signals from crystal. The intensities of $^{133}$Cs, $^{140}$Ce, $^{82}$Se increase dramatically while the intensity of $^{35}$Cl keeps unchanged, indicating the presence of $^{133}$Cs, $^{140}$Ce, $^{82}$Se and the absence of $^{35}$Cl in crystal. This conclusion could be further proved by comparing the individual element mass spectra with and without crystal. It is clearly illustrated in Fig.~\ref{Fig-ICP}(b)-(d) that strong peaks of $^{133}$Cs, $^{140}$Ce, $^{82}$Se in crystal can be identified, and no peak can be seen for the corresponding background. We note that LS-ICP-TOF technique is rather sensitive to the existing elements, for example, it can detect tiny amount of the elements from background. After zooming in the flat background in Fig.~\ref{Fig-ICP}(b), a weak peak from $^{133}$Cs can be observed. Similarly, a weak peak from $^{82}$Se can be identified in the background[Fig.~\ref{Fig-ICP}(d)]. No clear peak from $^{140}$Ce can be identified in the background [Fig.~\ref{Fig-ICP}(c)]. In addition, weak peaks of $^{35}$Cl from crystal and background can be identified in the signal. It is important to note that the $^{35}$Cl signals from crystal and background have almost identical intensity, which means that there is no extra Cl in the measured sample, except for the intensity from background. With these ICP analyses, we conclude that there is no Cl element in our CsCeSe$_2$ crystals.

To give the quantitative ratio of the elements, we performed new EDS measurements using Bruker energy-dispersive X-Ray Spectrometer equipped on a ZEISS EVO Scanning Electron Microscope (SEM) (EVO MA10). The EDS measurements also do not identify Cl element in the crystal. The elements ratio with EDS analyses is Cs : Ce : Se = 1 $\pm$  0.017 : 0.992 $\pm$ 0.020 : 1.996 $\pm$ 0.032, which is perfectly stoichiometric 1:1:2 within the (small) error bar.

\end{document}